\documentclass[conference,a4paper]{IEEEtran}

\newif\iffull
\fulltrue

\usepackage[hidelinks]{hyperref}
\usepackage[ruled,linesnumbered]{algorithm2e}
\usepackage[english]{babel}
\usepackage{amssymb}
\usepackage{amsmath}
\usepackage{xcolor}
\usepackage{cite}
\usepackage{graphicx}
\usepackage[utf8]{inputenc} 
\usepackage[T1]{fontenc}
\usepackage{url}
\usepackage{commath}
\usepackage{nicefrac}
\usepackage{complexity}
\usepackage{amsthm}
\usepackage{comment}
\usepackage{multirow}
\usepackage{enumitem}
\setlist[enumerate]{leftmargin=*}
\setlist[itemize]{leftmargin=*}

\usepackage[normalem]{ulem}

\usepackage{etoolbox}
\makeatletter
\patchcmd{\@makecaption}
  {\scshape}
  {}
  {}
  {}
\makeatletter
\patchcmd{\@makecaption}
  {\\}
  {.\ }
  {}
  {}
\makeatother

\theoremstyle{definition}
\newtheorem{theorem}{Theorem}
\newtheorem{lemma}{Lemma}
\newtheorem{claim}{Claim}
\newtheorem{definition}{Definition}
\newtheorem{corollary}{Corollary}
\newtheorem{problem}{Problem}
\newtheorem{construction}{Construction}
\newtheorem{example}{Example}

\newcommand{\bfg}{{\boldsymbol g}}

\renewcommand{\Bbb}{\mathbb}

\newcommand{\Z}{{\Bbb Z}}

\newcommand\bigzero{\makebox(0,0){\text{\huge0}}}
\newcommand{\minbinom}[2]{\lambda(#1,#2)}

\newcommand{\ceil}[1]{\left\lceil#1\right\rceil}
\newcommand{\floor}[1]{\left\lfloor#1\right\rfloor}

\usepackage{mleftright}       
\mleftright                   
\usepackage{booktabs}         

\begin{document}

\title{\textbf{Coding for IBLTs with Listing Guarantees}\vspace{-1.1ex}}
%

\author{%
   \IEEEauthorblockN{
                     \textbf{Daniella~Bar-Lev}\IEEEauthorrefmark{1}, 
                    \textbf{Avi~Mizrahi}\IEEEauthorrefmark{1},
                    \textbf{Tuvi~Etzion}\IEEEauthorrefmark{1},
                    \textbf{Ori~Rottenstreich}\IEEEauthorrefmark{1}, and
                    \textbf{Eitan~Yaakobi}\IEEEauthorrefmark{1}}
   \IEEEauthorblockA{\IEEEauthorrefmark{1}%
                     Department of Computer Science, 
                     Technion---Israel Institute of Technology, 
                     Haifa 3200003, Israel}
                      Email: \{daniellalev, etzion, avraham.m, or, yaakobi\}@cs.technion.ac.il
   \thanks{%
  This research was Funded by the European Union. Views and opinions expressed are however those of the author(s) only and do not necessarily reflect those of the European Union or European Research Council.}%
 }

 \IEEEoverridecommandlockouts

\maketitle

\begin{abstract}
The \emph{Invertible Bloom Lookup Table} (\emph{IBLT}) is a probabilistic data structure for set representation, with applications in network and traffic monitoring. It is known for its ability to list its elements, an operation that succeeds with high probability for sufficiently large table. However, listing can fail even for relatively small sets. This paper extends recent work on the worst-case analysis of IBLT, which guarantees successful listing for all sets of a certain size, by introducing more general IBLT schemes. These schemes allow for greater freedom in the implementation of the insert, delete, and listing operations and demonstrate that the IBLT memory can be reduced while still maintaining successful listing guarantees. The paper also explores the time-memory trade-off of these schemes, some of which are based on linear codes and \(B_h\)-sequences over finite fields.
\end{abstract}

\section{Introduction}
\label{sec:introduction}
The \emph{Invertible Bloom Lookup Table} (\emph{IBLT}) is a probabilistic data structure that is used for representing dynamic sets, with the ability to list the elements in the set~\cite{IBF, IBLT}. It has several applications, including traffic monitoring, error-correction codes for large data sets, and set reconciliation between two or more parties~\cite{li2016lossradar, li2016flowradar, BiffCodes, EppsteinGUV11, multiparty, Graphene19, kubjas2022partial, SetReconciliationWithPolinoms}. Although the listing operation of an IBLT might fail, it has been shown to be highly successful when the allocated memory is proportional to the number of elements~\cite{IBLT}. However, it is still possible to encounter failure in certain instances, such as when several elements are mapped to the same $k$ entries.

In this work, we consider the case of IBLT with 
listing guarantees in the worst-case.
Our point of departure here is a recent work
~\cite{IBLT_LFFZ} which introduced the problem of designing an IBLT with listing guarantees.
Under this setup, listing elements always succeeds for any set of up to $d$ elements from a finite universe.
They describe the mapping of elements to cells of the IBLT using a binary matrix, where each column represents an element from the universe and each row represents a cell of the IBLT.
Such a matrix is called \emph{$d$-decodable} if the listing is guaranteed to be successful for any set of up to $d$ elements.

In~\cite{IBLT_LFFZ}, the authors restricted the problem definition to only consider the so-called \emph{$d$-decodable matrices} and assumed that the element insertion, deletion, and listing operations as well as the structure of the IBLT were the same as the traditional IBLT (\autoref{fig:IBLT}).
This work extends the definition of a $d$-decodable matrix to a \emph{$d$-decodable scheme} by allowing greater freedom in the implementation of the insert, delete and listing operations and the design of the IBLT cells. Such an IBLT scheme, which is composed of a table, mapping matrix, and a set of operations, is called \emph{$d$-decodable} if its listing operation on its table is guaranteed to be successful whenever the number of elements in it is at most $d$. We show that such schemes can reduce the memory size of the IBLT. while still maintaining successful listing guarantees. The paper also explores the time-memory trade-off of these schemes, some of which are based on linear codes and $B_h$-sequences over finite fields.

This paper is organized as follows. \autoref{sec:defs} introduces the definitions that are used throughout the paper and the problem statements. \autoref{sec:decodable schemes} presents our constructions and lower bounds. Lastly, in \autoref{sec:time memory tradeoff} we discuss the time-memory trade-off for the different constructions. 

\section{Definitions and Problem Statement}
\label{sec:defs}
We start by formally defining IBLT schemes and its variants.
\begin{definition}
An \emph{\textbf{IBLT scheme}} consists of the following:
\begin{enumerate}
    \item A finite universe $U_n$ of size $n$ of all possible elements.
    \item A \emph{lookup table} $T$ which is a data structure that is composed of $m$ cells each of size $b$ bits. The size of the table $T$ is denoted by $s(T)=mb$. 
    \item An \emph{\textbf{IBLT protocol}} is a set of algorithms $P=(I, D, M, L)$ which are defined as follows:
    \begin{enumerate}
        \item \emph{Insert algorithm} ${I:U_n \times [m] \times \{0,1\}^{b} \rightarrow  \{0,1\}^{b}}$, which receives an element $u\in U_n$ (assuming it is not stored in the IBLT) and a cell state $\del{i, T_i} \in [m] \times \{0,1\}^{b}$ that contains the index and the content of the cell $T_i$. This algorithm updates $T_i$ to capture the insertion of $u$ into $T$.
        \item \emph{Delete algorithm} ${D\hspace{-2pt}:U_n \hspace{-1pt}\times\hspace{-1pt} [m] \hspace{-1pt}\times\hspace{-1pt} \{0,1\}^{b} \hspace{-2pt}\rightarrow \hspace{-2pt} \{0,1\}^{b}}$, which receives an element $u\in U_n$ (assuming it is stored in the IBLT) and a cell state $\del{i, T_i} \in [m] \times \{0,1\}^{b}$ that contains the index $i$ and the content of the cell $T_i$. This algorithm updates $T_i$ to capture the deletion of $u$ from $T$. 
        \item \emph{Mapping algorithm} $M:U_n \rightarrow 2^{[m]}$ which maps an element $u\in U_n$ to a subset of the cells in the lookup table (i.e., all cells that should be modified by the insertion/deletion of the element $u$).
        \item \emph{Listing algorithm} $L:\{0,1\}^{mb} \hspace{-2pt} \rightarrow \hspace{-2pt} 2^{U_n}$ which either lists all the elements that are stored in the IBLT, or fails.
    \end{enumerate}
\end{enumerate}
An IBLT scheme $(U_n, T, P)$ is called \textbf{\emph{$d$-decodable}} if it satisfies the \textbf{\emph{successful listing}} property:
if the lookup table $T$ stores a set $S \subseteq U_n$ of at most $d$ elements, then $L(T) = S$.
\end{definition}

\newcommand{\Lpeel}{L_{\text{peeling}}}
\newcommand{\ProtocolStandard}{P_s}
The definition of the IBLT protocol generalizes the well known definition of the traditional IBLT~\cite{IBF, IBLT}.
In our terminology, the traditional IBLT scheme, which we refer wherein as the standard scheme, can be described as follows.
\begin{definition}\label{def:standard scheme}
An IBLT scheme is called \textbf{\emph{standard}}, and is denoted by $(U_n, T_s, \ProtocolStandard=(I_s,D_s,M,\Lpeel))_S$, if the followings three conditions hold.
(1) The table $T_s$ is composed of $m$ cells, typically each of size $b=2\log{n}$. Each cell is composed of two components, a counter field, and a data field, which is referred as the \emph{xorSum field}.
(2) For $u \in U_n$, cell index $i\in[m]$, and cell state $T_i = {(t_\text{count}, t_\text{xorSum}) \in \cbr{0,1}^{\log n}\times \cbr{0,1}^{\log n}}$ we have that $I_s(u,i,T_i) = (t_\text{count} + 1, t_\text{xorSum}\oplus u)$ and $D_s(u,i,T_i) = (t_\text{count} - 1, t_\text{xorSum}\oplus u)$.
(3) $\Lpeel$ is a listing algorithm that operates as follows. First, it looks for a \emph{pure cell}, which is a cell whose counter is $1$, and deletes the corresponding element from the lookup table $T_s$ using the delete algorithm $D_s$. It continues this way until all elements are extracted from the lookup table or no pure cell exists. In the latter case, we say that the listing algorithm \emph{fails}.
\end{definition}

\begin{figure}[t]
    \centering
    \includegraphics[
     trim=0cm 0.8cm 0cm 0.5cm,
    width=.9\columnwidth,
    ]{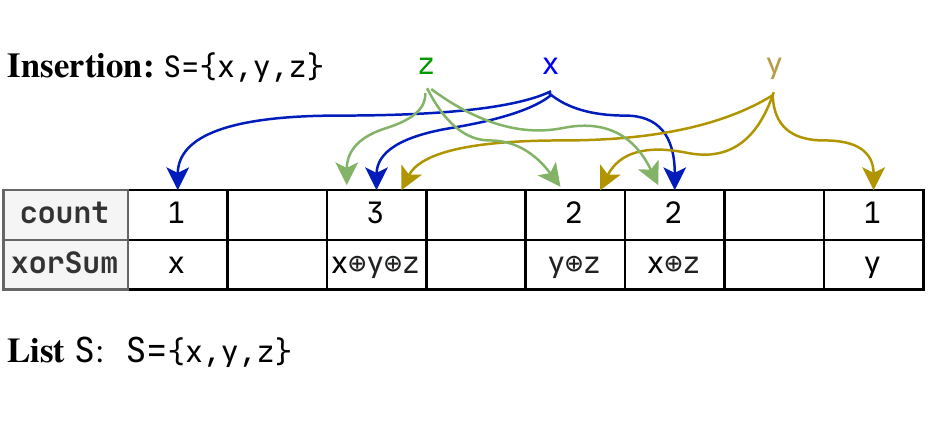}
    \caption{An IBLT representing the set $S=\{x,y,z\}$.
    Here the table $T_s$ consists of $m=8$ cells, each composed of a count field and a xorSum field.
    Element listing is possible starting from the most left cell for which the count field equals one, identifying $x$ as a member of $S$ and removing it from the other two cells. Then both $y$ and $z$ can be listed by the new pure cells.
    }
    \label{fig:IBLT}
\end{figure}

 In most works which studied IBLTs, the mapping $M$ in the standard scheme is implemented using a set of hash functions, each maps an element of $U_n$ to a subset of $[m]$. Here, $M(u)$ for $u\in U_n$ is defined to be the outputs of these functions.
An example of a standard IBLT scheme is depicted in \autoref{fig:IBLT}.

\begin{definition}
An IBLT scheme is called \textbf{\emph{standard-indel}}, denoted by $(U_n,\hspace{-0.3ex}T,\hspace{-0.3ex}P)_{\hspace{-0.3ex}SI}$, if its insert and delete algorithms are the same as in the standard schemes, i.e., $I\hspace{-0.5ex}=\hspace{-0.5ex}I_s$ and $D\hspace{-0.5ex}=\hspace{-0.5ex}D_s$.\footnote{This implies that every cell consists of a xorSum field and a counter field. However, we relax the assumption that the counter field is of size $\log n$. If necessary, the summation operations are calculated modulo the maximum number the counter can represent.}
\end{definition}

Hence, the main difference between a standard scheme and a standard-indel scheme is that while the former requires $\Lpeel$ to be the listing algorithm, this algorithm in the latter can be arbitrary. Furthermore, note that by definition, a standard IBLT scheme
implies a standard-indel IBLT scheme,
which implies an IBLT scheme.
In the cases where the IBLT scheme is a standard scheme 
or a standard indel scheme,
we describe the mapping using an $m\times n$ matrix $M$, in which $M_{i,j}= 1$ if and only if the $j$-th element of $U_n$ is mapped to the $i$-th cell of $T$.

\begin{example}\label{example:matrix}
    For the universe $U_6 = \{1, \ldots, 6\}$ consider the following binary matrix $M$.

    $$
    M=
    \begin{bmatrix}
    1 & 1 & 1 & 0 & 0 & 0\\
    0 & 0 & 0 & 1 & 1 & 1\\
    1 & 0 & 0 & 1 & 0 & 0\\
    0 & 1 & 0 & 0 & 1 & 0\\
    0 & 0 & 1 & 0 & 0 & 1\\
    \end{bmatrix}
     $$
    
    An IBLT based on $M$ has $m=5$ cells, each associated with a row of $M$.
    Such an IBLT, when containing for instance the set $S=\cbr{1, 3, 4}\subseteq[6]$, 
    has the counter array $\del{2, 1, 2, 0, 1}$ as the sum of the entries in the first, third, and fourth columns.
\end{example}

In this work we explore $d$-decodable IBLT schemes. Naturally, we seek to study the effect of the universe size $n$, the decodable threshold value $d$, and the protocol $P$ on the memory size used by the lookup table, $s(T)$. Additionally, we aim to find IBLT protocols that minimize the memory size $s(T)$. This problem can be formalized as follows.

\begin{problem}\label{problem:nd}
Given a set $U_n$, an integer $d \leq n$, find the values 
\begin{enumerate}
\item $s_{S}^*(n,d) = \min{\cbr{s(T): (U_n,T,P)_{S} \text{ is $d$-decodable}}}$,
\item $s_{SI}^*(n,d) = \min{\cbr{s(T): (U_n,T,P)_{SI} \text{ is $d$-decodable}}}$,
\item $s^*(n,d) = \min{\cbr{s(T): (U_n,T,P) \text{ is $d$-decodable}}}$. 
\end{enumerate}
\end{problem}

To limit the number of read or write memory accesses to a small fixed number when querying or inserting elements, we consider a specific family of IBLT schemes which we denote by $(d,k)$-decodable schemes. A $d$-decodable IBLT scheme is called \textbf{\emph{$(d,k)$-decodable}} if for any $u\in U_n$, its insert and delete algorithms affect exactly $k$ cells of $T$. This definition is extended also to the standard and standard-indel schemes. Similarly to \autoref{problem:nd}, we are interested in $(d,k)$-decodable IBLT schemes in which the size of the lookup table is minimal.

\begin{problem}\label{problem:ndk}
Given a set $U_n$, integers $d,k\leq n$, find the values

\begin{small}
\begin{enumerate}
\item $s_{S}^*(n,d,k) = \min{\cbr{s(T): (U_n,T,P)_{S} \text{ is $(d,k)$-decodable}}}$,
\item $s_{SI}^*(n,d,k) = \min{\cbr{s(T): (U_n,T,P)_{SI} \text{ is $(d,k)$-decodable}}}$,
\item $s^*(n,d,k) = \min{\cbr{s(T): (U_n,T,P) \text{ is $(d,k)$-decodable}}}$. 
\end{enumerate}
\end{small}
\end{problem}

We note that the results in this work demonstrate that the size of the lookup table $T$ can be significantly reduced by using general IBLT schemes as compared to the standard and standard-indel schemes. It is noteworthy to mention that this reduction in size may come with a trade-off in terms of increased computational time for the algorithms. A more detailed discussion of this trade-off is presented in \autoref{sec:time memory tradeoff}.
In the rest of this work, we assume that $b$ is at most $2\log(n)$. 

\begin{table*}
\centering
\caption{
\textbf{Constructions and lower bounds overview}. Note that $\Psi(n,d)$ is the maximum number of elements in a $B_d$-sequence in $GF(n)$ and $\minbinom{n}{k}$ is the minimal integer $m$ such that the number of unique length-$m$ binary vectors with weight $k$ is at least $n$.
}
\scriptsize
\begin{tabular}{c|cccccl}
\toprule
Scheme           & $d$         & $k$                 & $s^*$ - lower bound                   & $s^*$ - upper bound                                       & $b$                & Theorem\\
\hline
                 & $3$         & $-$                 &                                       & $2\ceil{\frac{3}{\log{3}}\log{n}}\ceil{\log{n}}$          & $2\ceil{\log{n}}$ & \cite{IBLT_LFFZ}\\
                 & $O(1)$      & $-$                 &                                       & $O\del{\log_2^{\floor{\log_2 d}+1}{n}}$                   & $2\ceil{\log{n}}$ & \cite{IBLT_LFFZ}\\
                 & $3$         & $2$                 & $2\ceil{2\sqrt{n}}\ceil{\log{n}}$     & $2\ceil{2\sqrt{n}}\ceil{\log{n}}$                         & $2\ceil{\log{n}}$ & \cite{IBLT_LFFZ}\\
standard         & $5$         & $3$                 &                                       & $6\del{2\ceil{\sqrt{n/6}}+1}\ceil{\log{n}}$               & $2\ceil{\log{n}}$ & \cite{IBLT_LFFZ}\\
\cite{IBLT_LFFZ} & $7$         & $4$                 &                                       & $8\ceil{\sqrt{2n}}\ceil{\log{n}}$                         & $2\ceil{\log{n}}$ & \cite{IBLT_LFFZ}\\
                 & $4$         & $2\ceil{\log(n+1)}$ &                                       & $8\ceil{\log(n+1)}\ceil{\log{n}}$                         & $2\ceil{\log{n}}$ & \cite{IBLT_LFFZ}\\
                 & any         & $d$                 &                                       & $2d \ceil{\sqrt{n}} \ceil{\log{n}}$                       & $2\ceil{\log{n}}$ & \cite{IBLT_LFFZ}\\
                 & any         & $O(d\log{n})$       &                                       & $O\del{d^2 \cdot \log^2{n}}$                              & $2\ceil{\log{n}}$ & \cite{IBLT_LFFZ}\\
\hline
                 & $3$         & $-$                 &                                       & $\del{\log{n}+2}\del{\log{n}+1}$                          & $\log{n}+2$       & \autoref{theorem:indel:expeel d3 d4}\\
standard-indel   & $4$         & $-$                 &                                       & $\del{\log(n+1) + 2}\del{2\log(n+1) + 1}$                 & $\log(n+1)+2$     & \autoref{theorem:indel:expeel d3 d4}\\
(this work)      & $3$         & $-$                 &                                       & $\del{\log{n}+1}^2$                                       & $\log{n}+1$       & \autoref{theorem:indel:improved d3}\\
                 & $3$         & any                 &                                       & $\del{\log{n}+1}\del{\minbinom{n}{k}+1}$                  & $\log{n}+1$       & \autoref{theorem:indel:improved d3}\\
\hline
                 & any         & $-$                 & $d\del{\log(n)-\log(d)}$              & $d\log(n+1)$                                              & $\log(n+1)$       & \autoref{theorem:general:d decodable lower bound},\autoref{theorem:general:d decodable}\\
general          & any         & $1$                 & $\ceil{\frac{n}{\Psi(n,d)-1}}\log(n)$ & $\ceil{\nicefrac{2n}{\del{\sqrt[d]{n}-2}}}\log(n)$        & $\log(n)$         & \autoref{theorem:general:bh:k=1 tight},\autoref{corollary:general:k=1}\\
(this work)      & $d>2$, even & $2$                 &                                       & $\del{\ceil{\nicefrac{2n}{(\sqrt[d]{n^2}-2)}}+1}\log(n)$  & $\log(n)$         & \autoref{theorem:general schemes:k=2}\\
                 & $4$         & $2$                 &                                       & $\del{2\ceil{\nicefrac{2n}{\del{3\sqrt{n}-4}}}+1}\log(n)$ & $\log(n)$         & \autoref{theorem:general schemes:d=4 k=2}\\
\bottomrule
\end{tabular}
\label{table:results}
\end{table*}

\section{Decodable Schemes}
\label{sec:decodable schemes}

So far in the literature only standard IBLT schemes were studied, while the other two variants are new to this work.

\subsection{Standard Schemes}
\label{sec:standard}
\newcommand{\mins} [2]{s^*_S\del{#1,#2}}    
\newcommand{\minsk}[3]{s^*_S\del{#1,#2,#3}} 

In this section we review previous results on standard decodable IBLT schemes, which are obtained by only selecting the mapping algorithm of the standard protocol. Each scheme is defined by an $m \times n$ binary mapping matrix $M$, and the size of its table is given by $s(T_s)=2m\log{n}$. \autoref{table:results} summarizes the relevant results from \cite[Table 1]{IBLT_LFFZ}. These are originally given as bounds on $m^*$, the minimal number of rows in a matrix $M$, which implies bounds on $s^*_S$. For more results and details the reader is referred to the original work~\cite{IBLT_LFFZ}.





    

    

\subsection{Standard-Indel Schemes}
\label{sec:indel}
This section considers the case of standard-indel schemes, i.e., schemes in which the insertion and deletion algorithms are as in the standard schemes. First, we note that to guarantee successful listing of up to $d$ elements, it is sufficient to use counters of size $\log{d}$, with modulo $d$ arithmetic.

\begin{claim}
It holds that $s^*_{SI}(n,d) \le \frac{\log{(n)} + \log{(d)}}{2\log{(n)}}\cdot s^*_{S}(n,d)$.
\end{claim}

\newcommand{\tcnt}{t_\text{count}}
\newcommand{\txsm}{t_\text{xorSum}}
\newcommand{\tmcnt}{T^{m}_\text{count}}
\newcommand{\tmxsm}{T^{m}_\text{xorSum}}

\newcommand{\Lexpeel}{L'_{\text{peeling}}}
Our main result in this section is that the memory size, $s(T)$, can be strictly reduced compared to the standard schemes by utilizing improved versions of the listing algorithm rather than $\Lpeel$.
As stated in \autoref{def:standard scheme}, the stopping condition of the peeling process is the absence of pure cells, i.e., there are non-empty cells, but none of them have a counter of 1. Next, we show that by adding a small step to this process, we can guarantee successful listing for small values of $d$. The key idea of this algorithm, denoted by $\Lexpeel$, is to overcome the absence of pure cells by searching for two cells which can reveal together the next element in the table. This is formally defined in \autoref{alg:extended peeling}, where the values of $\tcnt,\txsm$ in the $i$-th cell of $T$ is denoted by $T^{i}_\text{count}, T^{i}_\text{xorSum}$, respectively.

\setlength{\textfloatsep}{5pt}
\begin{algorithm}[!ht]
\caption{Extended peeling algorithm $\Lexpeel$}\label{alg:extended peeling}
    \KwIn{$d,T = \del{t_1, \dots, t_m}$}
    \KwOut{$S\subseteq U$, the elements that are stored in $T$}
    $J \gets \emptyset$\\
    \While{$T$ has a non-empty cell}{
        $s \gets \Lpeel(T)$\\

        \If{$s=\emptyset$}{
            \uIf{there are at most $d$ elements in $T$, and there exist $T_i,T_j$ such that $T^{i}_\text{count}-T^{j}_\text{count}=1$}{\label{alg:line:peeling fail}
                $e \gets T^{i}_\text{xorSum} \oplus T^{j}_\text{xorSum}$\\
                Delete $e$ from $T$\\
                $s \gets s \cup \cbr{e}$\\
            }\lElse{
                \Return{Failure}
            }
        }
        $J \gets J \cup s$\\
    }
    \Return{$J$}
\end{algorithm}

The advantages of the standard-indel schemes and $\Lexpeel$ over the standard schemes are presented next. 
\begin{theorem}\label{theorem:indel:expeel d3 d4}
The following holds.
\begin{enumerate}
    \item $s_{SI}^*(n,\hspace{-1pt}3) \leq \del{\log{n} + 2}\del{\log{n}+1}$ for $n=2^r$,
    \item $s_{SI}^*(n,\hspace{-1.5pt}4) \hspace{-1pt}\leq \hspace{-1.5pt}\del{\log(n\hspace{-1pt}+
    \hspace{-1.5pt}1) \hspace{-1.5pt}+\hspace{-1.5pt} 2}\hspace{-1.5pt}\del{2\log(n\hspace{-1pt}+\hspace{-1.5pt}1) \hspace{-1.5pt}+\hspace{-1.5pt} 1\hspace{-1pt}}$ for $n\hspace{-1.5pt}=\hspace{-1.5pt}2^r\hspace{-1.5pt}-\hspace{-1.5pt}1$.
\end{enumerate}
\end{theorem}
\begin{IEEEproof}
We start proving the second bound.
Let $M'$ be a binary parity check matrix for a length-$n$ linear code of with Hamming distance $5$, e.g. a parity check matrix for the binary Bose–Chaudhuri–Hocquenghem (BCH) code with $2\log(n+1)$ rows and $n$ columns\cite{roth_2006}.
Let $M$ be the matrix that is obtained from $M'$ by adding the all-ones row as the last row.
Let $T$ be the table with $m=2\log(n+1) + 1$ cells, each of size $b=\log(n+1)+2$. For each $i\in[m]$, we let $T_i = (t_\text{count}, t_\text{xorSum})\in\{0,1\}^2 \times \{0,1\}^{\log(n+1)}$ (i.e., the count field consists from two bits and the operations are done modulo $4$). Hence $s(T)=\del{\log(n+1) + 2}\del{2\log(n+1) + 1}$.
To prove the claim, we show that for the protocol $P=\del{I_s, D_s, M, \Lexpeel}$ we have that $(U_n, T, P)_{SI}$ is a $4$-decodable standard-indel scheme. Let $J \subseteq U_n$ be the set of elements in $T$, and assume $|J| \leq 4$.


Note that for a code of distance $d$, any set $C$ of at most $d-1$ columns of a parity check matrix is independent.
Let $C$ denote the set of columns in $M$ which corresponds with the elements of $J$, and observe that since $|C| \leq 4$ we have that the columns of $C$ are independent, and their sum must contain an entry with an odd value.
We show that $\Lexpeel$ always terminate successfully with the correct set $J$.
At any step in the loop where $\Lpeel$ fails (\autoref{alg:line:peeling fail}), there is no pure cell, which implies that $|J| \geq 3$.

If $|J|=3$, and since there is no pure cell, we have a cell $T_i$ for which $\tcnt=2$ and $\txsm=u \oplus v$ for $u,v\in J$.
Recall that the case where $|J|=3$ can be identified by $\tmcnt=3$, and $\tmxsm=u\oplus v \oplus w$ (where $J = \cbr{u,v,w}$).
Thus, $w= \txsm \oplus \tmxsm$ can be identified and deleted from $T$ and the algorithm can continue.
Otherwise, $|J|=4$, which can be identified since $\tmcnt=0$ and $T$ is not all zeroes.
As mentioned, the sum of the $4$ columns in $C$ must contain an odd value entry, and since there is no pure cell, the latter implies that there exist a cell $T_i$ with $\tcnt=3$.
By arguments similar to the ones for $|J|=3$, one element from $J$ can be correctly identified and removed and the algorithm can continue.

To prove the first bound, one can use similar arguments with the difference of letting $M'$ be the binary matrix that is composed of all the different $n$ columns of length $\log{n}$.
\end{IEEEproof}

Let $\minbinom{n}{k}\triangleq \min\cbr{m:\ n\geq \binom{m}{k}}$ be the minimal integer $m$ such that the number of unique length-$m$ binary vectors with weight $k$ is at least $n$. We can further improve the case where $d=3$ by cutting the cell counters width to one bit.

\begin{theorem}\label{theorem:indel:improved d3}
For $n=2^{r}$, we have that
${s_{SI}^*(n,3) \hspace{-1pt}\leq\hspace{-1pt} \del{\log{n}\hspace{-2pt}+\hspace{-2pt}1}^2}$,
and $
s_{SI}^*(n,3,k) \leq \del{\log{n}+1}\del{\minbinom{n}{k}+1}
$, for any $k\ge 1$.
\end{theorem}

The proof of \autoref{theorem:indel:improved d3} is similar to \autoref{theorem:indel:expeel d3 d4}, with the key idea that we can still identify the number of elements in the cells by utilizing the fields of the last cell.

\begin{IEEEproof}
Let $M$ be the binary $(\log (n) + 1) \times n$ matrix which is defined as follows. The first $m-1 = \log(n)$ rows of $M$ are composed of all the distinct binary column vector of length $\log(n)$ and the last row in $M$ is the all-one row.
Additionally, we let $T$ be a table with $m=\log(n)+1$ cells of size $b=\log(n)+1$ each. For each $i\in[m]$, we let $T_i = (t_\text{count}, t_\text{xorSum})\in\{0,1\} \times \{0,1\}^{\log n}$ (i.e., the count field consists from a single bit and the operations are done modulo $2$). Note that $s(T)=(\log(n)+1)^2$.
To prove the claim, we show that there exists a protocol $P=\del{I_s, D_s, M, L}$ such that $(U_n, T, P)_{SI}$ is a a $3$-decodable standard-indel scheme.

We denote by $\tmcnt, \tmxsm$ the value of $\tcnt,\txsm$ in the $m$-th cell of $T$, respectively (i.e, the cell that corresponds with the all-one row in $M$). We prove the claim by considering all the distinct cases based on the values of $\tmcnt$ and $\tmxsm$. Let $J\subset U_n$ be the set of elements in $T$ and assume $|J|\le 3$.
\begin{enumerate}
    \item If $\tmcnt=0$ and $\tmxsm=0$, then $\tmcnt=0$ implies that $|J|\in\{0,2\}$ and it holds that $J=\emptyset$ since otherwise there are exactly $2$ different elements $u,v$ in $T$. However, $\tmxsm = u \oplus v = 0$ and $u=v$, a contradiction.
    \item If $\tmcnt=0$ and $\tmxsm\ne 0$, then similar to the latter case, we have that $|J|=2$. Note that in this case $\tmxsm = u \oplus v$ and since $u\ne v$ the corresponding columns of $M$ are different in at least one row which guarantees that the corresponding cell in $T$ is pure. let $u$ denote the xorSum field in that cell, than $J=\{u, u\oplus\tmcnt\}$.
    \item If $\tmcnt=1$ and $\tmxsm=0$ then the size of $J$ must be odd (i.e., $|J|\in\{1,3\}$) and since $\tmxsm=0$, it must be that $J=\cbr{u,v,w}$. Note that since the three columns are all different, it can not be that the xorSum field of all the non zero cells of $T$ are the same. Hence, either we have a pure cell that contains a single element or a cell that contains exactly two elements. The former can be identified since its counter will be one and its xorSum field will be different than $\tmxsm$,
    while the latter can be identified since its counter will be zero and its xorSum field will be non-zero.
    If we have a pure cell $t$, we delete its element $\txsm$ .
    Otherwise, without loss of generality there is a cell $t$ with $\txsm = u \oplus v$, and we can reveal the third element by $w=\tmxsm\oplus\txsm$ and delete it from the table.
    In both cases we fall back to the previous case of $|J|=2$.
\end{enumerate}
This implies a listing algorithm $L$, and a standard indel scheme $(U_n,T,P)_{SI}$.
The proof for the second bound is similar apart from $M$ being an $m \times n$ binary matrix that composed of $m-1$ rows are all the distinct $\binom{m-1}{k}$ columns vectors of $m-1$ bits with exactly $k$ ones, and the last is again the all-ones row.
\end{IEEEproof}

    

\subsection{General Schemes}
\label{sec:general}
In this section we consider general IBLT schemes (i.e., not standard or standard indel schemes). That is, here we allow the modification of the insert and delete algorithms as well as the listing and mapping algorithms and the design of the lookup table $T$. More precisely, we allow the insert and delete algorithms to rely on computations over finite fields. For an integer $r\ge 1$, let $GF(2^r)$ be the Galois Field of size $2^r$. 
In the rest of this section all the operations are the field operations. 
Proofs for several of the claims are given in this section, while the rest can be found in the appendix.

Define $T_{F}$ to be the lookup table that consists of $m$ cells, each of size $r$ bits. For an $m\times n$ matrix $H\in GF(2^r)^{m\times n}$, let $M_{H}$ be the mapping algorithm that maps the $j$-th element of $U_n$ to the $i$-th cell of $T_{F}$ if and only if $H_{i,j}\ne 0$. Additionally, let $I_H$ and $D_H$ be the insert and delete algorithms which are defined as follows. For $0\le i\le d$ and $u_j\in U_n$, $j\in [n]$, let $(i,T_i)$ be the current cell state of the $i$-th cell of $T$. Then 
    \begin{align*}
    I_H(u_j,i,T_i) \triangleq T_i + H_{i,j},\ \ \ \text{and} \ \ \ 
    D_H(u_j,i,T_i) \triangleq T_i + H_{i,j}.
    \end{align*}
    
Note that if we consider $T_{F}$ as a vector of length $m$ over $GF(2^r)$, then the state of $T_{F}$ after inserting or deleting an element $u_j\in U_n$ is equal to $T_{F}+h_j$, where $h_j$ is the $j$-th column of $H$. Hence, since the characteristics of the field is $2$, if the lookup table stores a set $S\subseteq U_n$ then the state of $T_{F}$ is given by $T_{F}(S)=H\cdot v_S$, where $v_S\in\{0,1\}^n$ is the indicator binary length-$n$ vector $n$ such that $\mathsf{supp}(v_S) = S$  (i.e., the $j$-th entry of $v_S$ is one if and only if $j\in S$). 

We start by showing that the size of the required lookup table $T_F$ can be drastically reduced using $I_H,D_H$ by selecting a suitable matrix $H$. To this end, first consider the following $d\times n$ matrix, for $n=2^r-1$, where $\alpha$ is primitive in $GF(2^r)$.

\begin{small}
$$
H^d_{n, \alpha} \hspace{-3pt}= \hspace{-2pt}\left[\begin{matrix}
1 & \alpha & \alpha^2 & \alpha^3 & \cdots & \alpha^{n-1}\\
1 & \alpha^3 &  \alpha^{3\cdot2} & \alpha^{3\cdot3} & \cdots &  \alpha^{3(n-1)}\\
1 & \alpha^5 &  \alpha^{5\cdot2} &  \alpha^{5\cdot3} & \cdots &  \alpha^{5(n-1)}\\
\vdots & \vdots & \vdots & \vdots & \ddots & \vdots\\
1 & \alpha^{2d-1} &  \alpha^{(2d-1)2} &  \alpha^{(2d-1)3} & \cdots &  \alpha^{(2d-1)(n-1)}
\end{matrix}\right]
$$
\end{small}

It is well known that $H^d_{n, \alpha}$ is a parity-check matrix of a primitive narrow-sense BCH code~\cite{roth_2006} with minimum distance ${t\ge 2d+1}$.  That is, $H^d_{n, \alpha}$ is a parity-check matrix of a $d$-error-correcting-code of length $n=2^r-1$. 

\begin{theorem}
\label{theorem:general:d decodable}
 For $n=2^r-1$ and $d\ge 1$, it holds that ${s^*(n, d)\le d \cdot r = d\log(n+1).}$ 
\end{theorem}

\begin{IEEEproof}
To prove the result, we show that for $H=H_{n,\alpha}^d$, there exists a listing algorithm $L$ such that the scheme $(U_n, T_{F}, P=(I_H,D_H,M_H,L))$ is a $d$-decodable IBLT scheme. Since $s(T_{F}) = d\log (n+1)$ the result follows. 
    
Note that $H$ is a parity-check matrix of a code that corrects $d$ errors. Hence, for any $S\subseteq U_n$ such that $|S|\le d$, the syndromes $H\cdot v_S$ are unique. That is, for any $|S|\le d$, the state of $T_{F}$ is unique which implies that there exists a listing algorithm $L$ for which $(U_n, T_{F}, P=(I_H,D_H,M_H,L))$ is $d$-decodable. 
\end{IEEEproof}

The listing algorithm for the protocol in the latter proof can be implemented using a decoder for the BCH code. Some of the known decoders use the Berlekamp-Massey algorithm, which is very efficient.
In~\cite{schipani2011decoding}, the authors presented an efficient decoder for large values of $n$ with $\max\cbr{O(d\sqrt{n}), O(d^2\log(n)}$ time complexity. 
We note that the rest of this section considers mappings that are based on the parity-check matrix of the BCH code, and that the BCH decoders can be used for listing in these cases as well by introducing some adjustments with low time overhead.

Before we consider the case of 
$(d,k)$-decodable IBLT schemes, we present a lower bound on $s^*(n,d)$ in the following lemma.

\begin{lemma}
\label{theorem:general:d decodable lower bound}
It holds that 
$
s^*(n,d)
> d\log(n) - d\log(d).
$
\end{lemma}

\begin{IEEEproof}
    Let $(U_n,T,P)$ be a $d$-decodable IBLT scheme. By definition, the listing algorithm $L$ is guaranteed to successfully list any set $S\subseteq U_n$ of size at most $d$ which is stored in $T$. That is, the state of $T$ is unique for any such set $S$. Hence, we have that $2^{s(T)}\ge \sum_{i=0}^d \binom{n}{i}$ which implies that  
    \begin{align*}
    s^*(n,d)& \ge \log\left(\sum_{i=0}^d \binom{n}{i}\right) > \log\binom{n}{d}\\
    & \ge \log\left(\left(\frac{n}{d}\right)^d\right) = d\log(n) - d\log(d).
    \end{align*}
\end{IEEEproof}

\begin{corollary}\label{corollary:general:s=theta dlogn}
    If $d=O(\polylog (n))$ then 
    $
    \lim_{n\to\infty}\frac{s^*(n,d)}{d\log(n)} = 1.
    $
\end{corollary}

\begin{IEEEproof}
    From \autoref{theorem:general:d decodable} and \autoref{theorem:general:d decodable lower bound} we have that 
    \[
    1 - \frac{\log(d)}{\log(n)} < \frac{s^*(n,d)}{d\log (n)} \le \frac{\log(n+1)}{\log(n)}.
    \]
    Since $d=O(\polylog(n))$ we have that $\lim_{n\to\infty}\frac{\log(d)}{\log(n)} = 0$ which completes the proof. 
\end{IEEEproof}

Next we discuss IBLT schemes which are $(d,k)$-decodable. Clearly, for $k=d$ the $d$-decodable scheme that is presented in the proof of \autoref{theorem:general:d decodable} is a $(d,k)$-decodable scheme which implies the following result. 

\begin{corollary}\label{corollary:general:k ge d}
For $n=2^r-1$ and $d\ge 1$, we have that ${s^*(n, d, k\ge d,P)\le k\log(n+1).}$ \footnote{Note that if $k>d$ then we can append $k-d$ redundant rows to $M_{n,\alpha}^d$ and use a similar construction to obtain that $s^*(n, d, k)\le k\log_2(n+1)$.} 
\end{corollary}
 To discuss the more involved case of $k<d$, we first give the definition of $B_h$-sequences~\cite[Section V]{brouwer2006new},\cite{bose1960theorems,o2004complete}.
 
\begin{definition}
    A sequence $g_1,\ldots,g_n$ in an Abelian group $G$ is called a \textbf{\emph{$B_h$-sequence}} if all the non-zero sums $g_{i_1}+g_{i_2}+\cdots+g_{i_h}$, for $1\le i_1 \le i_2 \le \cdots \le i_h\le n$ are distinct in $G$.
\end{definition}

It can be readily verified that any $B_h$-sequence is also a $B_{h'}$-sequence for $1\le h'\le h$. The following is an example for a $B_2$-sequence. 

\begin{example}
Let $G=(\Z, + )$ and consider the sequence $1,2,5,7$. It can be verified that all the sums $a+b$ for $a,b\in\{1,2,5,7\}$ are distinct and hence $1,2,5,7$ is a $B_2$-sequence. Since we have that $1+1+7=2+2+5=9$, the latter sequence is not a $B_3$-sequence. 
\end{example}

In this work, we only consider $B_h$-sequences in $GF(2^r)$. Denote by $\Psi(2^r,h)$ the maximum number of elements in such a $B_h$-sequence.
In the next theorem, we present an upper bound on $s^*(n,d,k=1)$ under the assumption that the size of each cell in a lookup table $T_{F}$ is $b=\log (n)$.

\begin{theorem}\label{theorem:general:bh:k=1}
    Let $n=2^r$. If the size of each cell in $T$ is $b=\log (n)$, then 
    $
    s^*(n,d,k=1)\le \ceil{\frac{n}{\Psi(n,d)-1}}\log (n).
    $
\end{theorem}

\begin{IEEEproof}
    Let $\ell\triangleq \Psi(n,d)-1$ and let $0,g_1,\ldots, g_{\ell}$ be a ${B_d\text{-sequence}}$ of length $\ell+1$ in $GF(2^r)$. 
     We define the row vector $\bfg=(g_1,g_2,\ldots,g_\ell)$.
Let ${m=\ceil{\frac{n}{\Psi(n,d)-1}}}$ and let us show that there exists a protocol $P$ such the $(U_n, T, P)$ is $(d,1)$-decodable and $T=T_F$ is a table with $m$ cells. Let $H_{\bfg}$ be the following $m\times n$ diagonal block matrix,  

    $$
    H_{\bfg}=\left[
    \begin{matrix}
        \bfg &          &          &          & \\
             & \bfg     &          & \bigzero & \\
             &          &  \ddots  &          & \\
             & \bigzero &          & \bfg     & \\
             &          &          &          & \bfg'
    \end{matrix}
    \right]
    $$

    where $\bfg'$ is a shortening of $\bfg$ to its first entries such that the matrix $H$ has $n$ columns.
    For $I=I_H,D=D_H, M=M_H$, if $T_F$ contains a subset $S\subseteq[n]$ of size at most $d$, then by the definition of the insertion, deletion algorithms, each cell of $T_F$ contains the sum of at most $d$ different elements from a $B_d$-sequence. Hence, for any such $S$, the state of $T_F$ is unique and there exists a listing algorithm $L$ such that  $(U_n, T_F, P=(I_H,D_H,M_H,L))$ is $(d,1)$-decodable. 
\end{IEEEproof}

Next we show that the bound in \autoref{theorem:general:bh:k=1} is tight.

\begin{theorem}\label{theorem:general:bh:k=1 tight}
    Let $n=2^r$. If the size of a each cell in $T$ is $b=\log(n)$, then 
    $
    s^*(n,d,k=1)= \ceil{\frac{n}{\Psi(n,d)-1}}\log(n).
    $
\end{theorem}

\begin{IEEEproof}
    By \autoref{theorem:general:bh:k=1} we have that $s^*(n,d,1)\le \ceil{\frac{n}{\Psi(n,d)-1}}\log(n)$. Assume to the contrary that there exists a $d$-decodable IBLT scheme $(U_n,T,P)$ for which $s(T)< \ceil{\frac{n}{\Psi(n,d)-1}}\log(n)$. That is, the number of cells in $T$ is 
    $m< \ceil{\frac{n}{\Psi(n,d)-1}}$. By the pigeonhole principle, the latter implies that there exists a cell $j\in[m]$ such that $j\in M(u)$ for at least $\Psi(n,d)$ elements $u\in U_n$. 
    Denote $U = \left\{u\in U_n:\ j\in M(u)\right\}$. Since $0\notin U$, the set $U\cup\{0\}$ is a $B_d$-sequence with more than $\Psi(n,d)$, which results with a contradiction.
\end{IEEEproof}

To conclude the discussion regarding $(d,1)$-decodable IBLT schemes, we present a construction of such a $B_d$-sequence in $GF(2^r)$~\cite[Section V]{brouwer2006new}.

\begin{construction}\label{construction: Bd-sequence}
For $n=2^r$ and an integer $d$, let $n'+1$ be the largest power of $2$ such that $d\log (n'+1)\le r$. Additionally, let $H_{n'}^d$ be a BCH parity-check matrix with distance at least ${2d+1}$ over $GF(n'+1)$. It can be verified that the sequence $g_0=0,g_1,\ldots, g_{n'}$, such that for $1\le i\le n'$, $g_i = h_i$ is the $i$-th column of $H_{n'}^d$ is a $B_d$-sequence in $GF(n)$. Therefore, $\Psi(n,d) \geq n'+1\geq \sqrt[d]{n}/2$.
\end{construction}

Using \autoref{construction: Bd-sequence}, $H_\bfg$ is a matrix with $m\hspace{-0.6ex} =\hspace{-0.6ex} \ceil{\hspace{-0.6ex}\frac{2n}{\sqrt[d]{n}-2}\hspace{-0.6ex}}$ rows, and thus we get the following upper bound on $s^*(n,d,k=1)$.

\begin{corollary}\label{corollary:general:k=1}
    Let $n=2^r$. If the size of each cell in $T$ is $b=\log(n)$ then
    $
     s^*(n,d,k=1) \le\ceil{\frac{2n}{\sqrt[d]{n}-2}}\log(n).
    $
\end{corollary}

Next we address the more intriguing case of $k=2$. Due to space limitations, we only discuss the case where $d$ is even. Note that for odd $d$, the results for $d+1$ are applicable, however, they are probably not optimal. 

\begin{construction}\label{construction: Bd for k=2}
    For $n=2^r$ and an even integer $d$, let $n'+1$ be the largest power of $2$ such that ${d\log(n'+1)}\le 2r$. Additionally, let $H_{n'}^d$ be a BCH parity-check matrix with distance at least ${2d+1}$ over $GF(n')$ and denote by $H^{(U)}$, $H^{(L)}$ the upper, lower half of $H_{n'}^d$, respectively. As in \autoref{construction: Bd-sequence}, let $\bfg^{(U)},\bfg^{(L)}$ be the vector of length $n'$ such that $g_i^{(U)},g_i^{(L)}$, is equal to the $i$-th column of $H^{(U)}$, $H^{(L)}$, respectively. Define $H_2$ as the following block matrix 
    $$
    H_2 = \left[
    \begin{matrix}
        \bfg^{(U)}  &            &           &          &           &\\
        \bfg^{(L)}  & \bfg^{(U)} &           &          &\bigzero   &\\
                    & \bfg^{(L)} & \bfg^{(U)}&          &           &\\
                    &            & \ddots    & \ddots   &           &\\
                    &\bigzero    &           &\bfg^{(L)}& \bfg^{(U)}&\\
                    &            &           &          &\bfg^{(L)} &\bfg^{(U)}\\
                    &            &            &           &         &\bfg^{(L)} 
    \end{matrix}
    \right].
    $$
\end{construction}

Using $H_2$ from \autoref{construction: Bd for k=2}, we have the following result.

\begin{theorem}\label{theorem:general schemes:k=2}
    Let $n=2^r$. If the size of each cell in the lookup table $T$ is $b=\log(n)$, then for any even integer $d>2$ we have that 
    $
    s^*(n,d,k=2) \le \left(\ceil{\frac{2n}{\sqrt[d]{n^2} -2}}+1\right)\log(n).
    $
\end{theorem}

\begin{IEEEproof}
For $H=H_2$, let $I=I_H,D=D_H$ and $M=M_H$. To prove the claim, we show that the sum of any $d$ or less columns is unique. Let us divide the columns of $H$ to $m-1$ disjoint sets $C_1,C_2,\ldots, C_{m-1}$ according to the location of their first non-zero entry. 

First note that if all the columns are taken from the same $C_i$, similarly to the proof of \autoref{theorem:general:d decodable}, their sum is unique. 
Otherwise, assume to the contrary that there exist two sums of up to $d$ columns which are the same. Since the characteristic of the field is $2$, the latter implies that there exists a selection of up to $2d$ columns in $H$ such that their sum is zero, and denote this set of columns by $C$. Let $i$ be minimal such that $C\cap C_i\ne\emptyset$ and note that the $i$-th entry in the sum of the columns in $C\cap C_i\ne\emptyset$ must be zero. However, since the elements are taken from $\bfg^{(U)}$, which is a $B_{d/2}$-sequence, the latter implies that $|C\cap C_i|>d$. By the same arguments we also have that $|C\cap C_{i'}|>d$, where $i'$ is the maximal integer for which $C\cap C_{i'}\ne\emptyset$. Thus, by recalling that $C_i\cap C_{i'}=\emptyset$, we have that $|C|>2d$, a contradiction.
\end{IEEEproof}

In the rest of this section, we deal with the special case of $d=4$ and show that the bound in \autoref{theorem:general schemes:k=2} can be improved using a more sophisticated selection of the matrix $H$.

\begin{construction}
    \label{construction: Bd for k=2 d=2 improved}
    For $n=2^r$ let $n'+1$ be the largest power of $2$ such that $2\log(n'+1)\le r$. Denote by $\widehat{H}_{n'}^4$ the matrix that is obtained from  $H_{n'}^4$ by removing its left-most column. Similarly to \autoref{construction: Bd for k=2}, let $H^{(U)},H^{(L)}$ be the lower, upper half of $\widehat{H}_{n'}^4$ and let $\bfg^{(U)},\bfg^{(L)}$ be the length-$(n'-1)$ vector representation of $H^{(U)},H^{(L)}$, respectively. Define 

    $$
    \widehat{G_2}= \left[
    \begin{matrix}
    \bfg^{(U)} &  {\bf 0}    & \bfg^{(U)} \\
    \bfg^{(L)} &\bfg^{(U)}   &{\bf 0}     \\
     {\bf 0}   &\bfg^{(L)}   & \bfg^{(L)}
    \end{matrix}\right],
    $$
    
    \hspace{-3ex} and let $\bfg_1,\bfg_2,\bfg_3$ denote the three rows of $\widehat{G_2}$. We let $\widehat{H_2}$ be the following matrix 
    
    $$
    \widehat{H_2} = \left[
    \begin{matrix}
       \bfg_1  &         &        &         &\\ 
       \bfg_2  &         &        &         &\\
       \bfg_3  & \bfg_1  &        &         &\\
               & \bfg_2  &        &         &\\
               & \bfg_3  & \bfg_1 &         &\\
               &         & \ddots & \ddots  &\\
               &         &        & \bfg_3  & \bfg_1\\
               &         &        &         & \bfg_2\\
               &         &        &         & \bfg_3
    \end{matrix}
    \right].
     $$
\end{construction}

Using \autoref{construction: Bd for k=2 d=2 improved}, we have the following result.

\begin{theorem}\label{theorem:general schemes:d=4 k=2}
     Let $n=2^r$. If the size of each cell in the lookup table $T$ is $b=\log(n)$, then we have that 

    $$
    s^*(n,d=4,k=2) \le \left(2 \ceil{\frac{2n}{3\sqrt{n}-4} } + 1\right)\log(n).
    $$
\end{theorem}

\begin{IEEEproof}
 Let $H=\widehat{H_2}$, similar to the proof of \autoref{theorem:general schemes:k=2}, it is sufficient to show that the sum of any $j\le 8$ columns in $H$ is non zero. Here we only show that the sum of any $j\le 8$ columns in $G=\widehat{G_2}$ is non-zero, while the generalization for $H$ is similar and can be done using the same technique. Divide the columns of $G$ to three disjoint sets $C_1,C_2,C_3$ according to their non-zero entries such that $C_1$ is the set of the lest-most columns and $C_3$ is the set of the right-most columns. Let $C$ be a set of $j\le 8$ columns from $G$ and consider the following cases.
 \begin{enumerate}
     \item If all the columns of $C$ belong to the same set $C_i$, then their sum can not be zero since the concatenation of $\bfg^{(U)}$ and $\bfg^{(L)}$ forms the sub-matrix that is obtained by a single column removal from the parity-check matrix $H_{n'}^4$ of a code with distance $9$.
     \item If all the columns of $C$ belongs to exactly two sets, $C_i$ and $C_j$, then by the pigeonhole principle, we have w.l.o.g.  $C' = |C\cap C_i| \le 4$. Since all the columns from $C_j$ have the a zero entry in the same row, the sum of all the columns in $C$ and the sum of the columns in $C'$ are the same at this row. Hence, the value in this entry is the sum of $1\le j'\le 4$ elements from either $\bfg^{(U)}$ or $\bfg^{(L)}$. Note that both $\bfg^{(U)}$ and $\bfg^{(L)}$ are $B_2$-sequences and hence the sum in the latter entry can not be zero. 
     \item Otherwise $C$ contains columns from each of the sets $C_1,C_2,C_3$. Assume to the contrary that the sum of the columns in $C$ is zero and order the element of $C_i$, $i\in[3]$ by theire order in $G$. It can be verified that for any $i\in[3]$, if $C$ contains the $\ell$-th columns of $C_i$, in order for the first and last entries of the sum to be zero, $C$ must also contain the $\ell$-th column of the other two sets. Hence, we have that $|C|\in\{3,6\}$.
     In both cases, by writing the simple algebraic equations it can be verified that the sum can not be zero, a contradiction.
 \end{enumerate}
\end{IEEEproof}

\section{Time vs Memory Trade-Off}\label{sec:time memory tradeoff}

The results in this work demonstrate that successful listing can be achieved in the worse-case with IBLT size which is significantly smaller than the ones presented in \cite{IBLT_LFFZ}. This improvement is the result of our relaxation of the IBLT scheme to allow the use of arbitrary algorithms. It is also worth mentioning that some of the standard schemes from \cite{IBLT_LFFZ} require either additional memory to store the mapping matrix or additional time to calculate the corresponding columns during the insert and delete operations.
On the contrary, for all of our constructions the only additional memory that might be needed is for storing the primitive element $\alpha$ ($\log(n)$ bits) and the computation of the relevant column can be done in $O(d\log^2(n))$, which is slightly higher than the probabilistic IBLT implementations~\cite{IBLT}.

The time complexity of the listing operation of the standard and standard-indel schemes is the same as that of the probabilistic IBLT, i.e., $O(m)$ where $m$ is the number of cells in the lookup table.
However, the listing time of the constructions in \autoref{sec:general} is $\max\cbr{O(d\sqrt{n}), O(d^2\log(n)}$.

\bibliographystyle{IEEEtran}
\bibliography{IBLTcodes}






\end{document}